# Time Delay Plot for $\pi N$ Elastic Scattering

**Mohamed E. Kelabi**
Physics Department, University of Tripoli, Tripoli, LIBYA.

**Abstract**
We evaluated the time delay plot in the established region of the Δ(1232) resonance through the use of $\pi N$ elastic scattering phase shift analysis of the partial wave amplitude $P_{33}$. The pole position and width of the Δ(1232) resonance were obtained and found in agreement with earlier calculations.

## Introduction

The definition of resonance has been a matter of much debate [1]. Different approaches are used to identify resonance and determining its parameters e.g., using partial wave analysis of meson-paryon scattering data then finding the poles of $T$-Matrix or using techniques such as Argand diagrams and speed plot [2], [3]. In contrast of these conventional procedures we make use of one of the basic criteria for the existence of resonance by means of positive peak in the time delay. Intuitively one would expect the scattering particles to be held up for a while during the formation and decay of a resonance leading to a positive time delay peak in energy around the resonance mass. This time delay peak is related to the lifetime of a resonance. It was noted [4] that a bump in cross section may not always be due to a resonance. On the other hand, it has been repeatedly mentioned in many literatures as well as textbooks [5], [6], [7] that a positive maximum in time delay plot is a necessary condition for the existence of a resonance.

## Time delay and resonance

The concept of time delay $\Delta t$ was originally introduced by Eisenbud [8] and discussed by Wigner [9], Dalitz and Moorehouse [1], Bohm [10], and Nussenzveig [11]. It is defined as the difference between the time in which a wave packet passes through an interaction region and the time spent by a free wave packet passing through the same distance. In the early fifties it was shown [8], [9], [10] using the wave packet analysis that the time delay in collisions can be defined in terms of energy derivative of the scattering phase shift. For a single channel elastic scattering, the time delay of the outgoing wave packet with respect to the non-interacting wave packet is given by [9], [11]

$$\Delta t(W) = \mathrm{Re}\left(-i\,\frac{1}{S(W)}\frac{dS(W)}{dW}\right) \tag{1}$$

in natural units [12]. The scattering matrix $S$ is related to the scattering phase shift $\delta$ by the relation

$$S(W) = e^{2i\delta(W)} \tag{2}$$

where $W$ is the total center of mass energy of $\pi N$ system. Eqs. (1) and (2) lead to a simple expression of the time delay

$$\Delta t(W) = 2\frac{d\delta(W)}{dW} \tag{3}$$

This result is larger than the result of Eisenbud [8] by a factor of 2 [13]. A reason for this factor was noted by Wigner [14], [15]. At high energies, in addition to elastic scattering, the possibility of scattering into some inelastic channels also takes place. In this case the elastic *S*-matrix element becomes [16]

$$S(W) = \eta e^{2i\delta(W)} \tag{4}$$

where $\eta$ is the inelasticity parameter defined such that $0 < \eta \leq 1$. Substituting Eq. (4) into Eq.(1), gives

$$\Delta t(W) = \text{Re}\left[-i\left(2i\frac{d\delta(W)}{dW} + \frac{1}{\eta}\frac{d\eta}{dW}\right)\right] = 2\frac{d\delta(W)}{dW} \tag{5}$$

Thus the time delay for elastic scattering is the same, irrespective of the presence of inelastic channels. The scattering phase shift is still real, but its value is affected by the presence of the inelastic channels.

Instead of using the phase shift formulation of the *S*-matrix , one can start by using the transition matrix *T*, defined by the relation

$$S = 1 + 2iT \tag{6}$$

where the complex *T*-matrix

$$T = \text{Re}(T) + i\ \text{Im}(T) \tag{7}$$

containing the entire information of the resonant and non-resonant scattering. Substituting Eq. (6) into Eq. (1) gives the average time delay in terms of the real and imaginary parts of the amplitude *T* in the elastic scattering [17],

$$S^*S\,\Delta t = 2\left(\frac{d}{dW}\text{Re}(T) + 2\,\text{Re}(T)\frac{d}{dW}\text{Im}(T) - 2\,\text{Im}(T)\frac{d}{dW}\text{Re}(T)\right) \tag{8}$$

where $S^*S$ can be evaluated using Eq. (6).

In the present work, we have evaluated the time delay of $\pi N$ scattering for the partial wave $P_{33}$. We start by writing Eq. (6) in the following form

$$T(W) = \frac{S(W) - 1}{i} \tag{9}$$

and substituting Eq. (2) in Eq. (9) obtaining

$$T(W) = \frac{1}{\cot\delta(W) - i} \tag{10}$$

By using Taylor expansion for $\cot\delta(W)$ about the resonance energy $W_r$

$$\cot\delta(W) = \cot\delta(W_r) + (W - W_r)\frac{d}{dW}\cot\delta(W)\Big|_{W=W_r} + \ \ldots \qquad (11)$$

providing that the resonance height is much large compared with the resonance width. At the resonance the phase shift passes through $\pi/2$, giving

$$\cot\delta(W_r) = 0$$

we define

$$\frac{2}{\Gamma} = -\frac{d}{dW}\cot\delta(W)$$

at the resonance, then Eq. (10) and Eq. (11), respectively, give the following forms

$$T(W) = \frac{\Gamma/2}{(W_r - W) - i\,\Gamma/2} \qquad (12)$$

$$\delta(W) = \cot^{-1}\left(\frac{W_r - W}{\Gamma/2}\right) \qquad (13)$$

The derivative of Eq. (13) gives the time delay $\Delta t(W)$,

$$2\frac{d\,\delta(W)}{dW} = \frac{\Gamma}{(W_r - W)^2 + (\Gamma/2)^2} \qquad (14)$$

The maximum of the time delay occurs at the resonance, $W = W_r$, giving

$$\Delta t(W_r) \equiv H = 4/\Gamma \qquad (15)$$

where $H$ is the height of the resonance.

**Calculations**

We employ an accurate parameterization of the phase shift $\delta$ ( $\delta_{33} = \delta_{IJ}$ ; $I = J = 3/2$) [18],

$$q^3(W)\cot\delta(W) = a_0 + a_2 q^2(W) + a_4 q^4(W) + a_6 q^6(W) \qquad (16)$$

where

$$q^2(W) = \frac{[W^2 - (m_N + m_\pi)^2][W^2 - (m_N - m_\pi)^2]}{4W^2} \qquad (17)$$

is the total CM momentum of $\pi N$ system, and $m_N$ , $m_\pi$ are the nucleon and pion masses, respectively. Using an updated run of $\pi N$ scattering data [19], for single channel, one can obtain the following fits

$$a_0 = 5.22322$$
$$a_2 = -0.92458$$
$$a_4 = -0.21266$$
$$a_6 = -0.06642$$

where we have used the masses $m_N = 6.723$ for the proton , and $m_\pi = 1$ for charged pion in Eq. (17).

In what follows we plot the result of our calculations: Figure 1 shows our fit of the scattering phase shift $\delta$, for partial wave $P_{33}$, compared with data, and in figure 2 we plot the result of our calculated transitional amplitude $T$ compared with data.

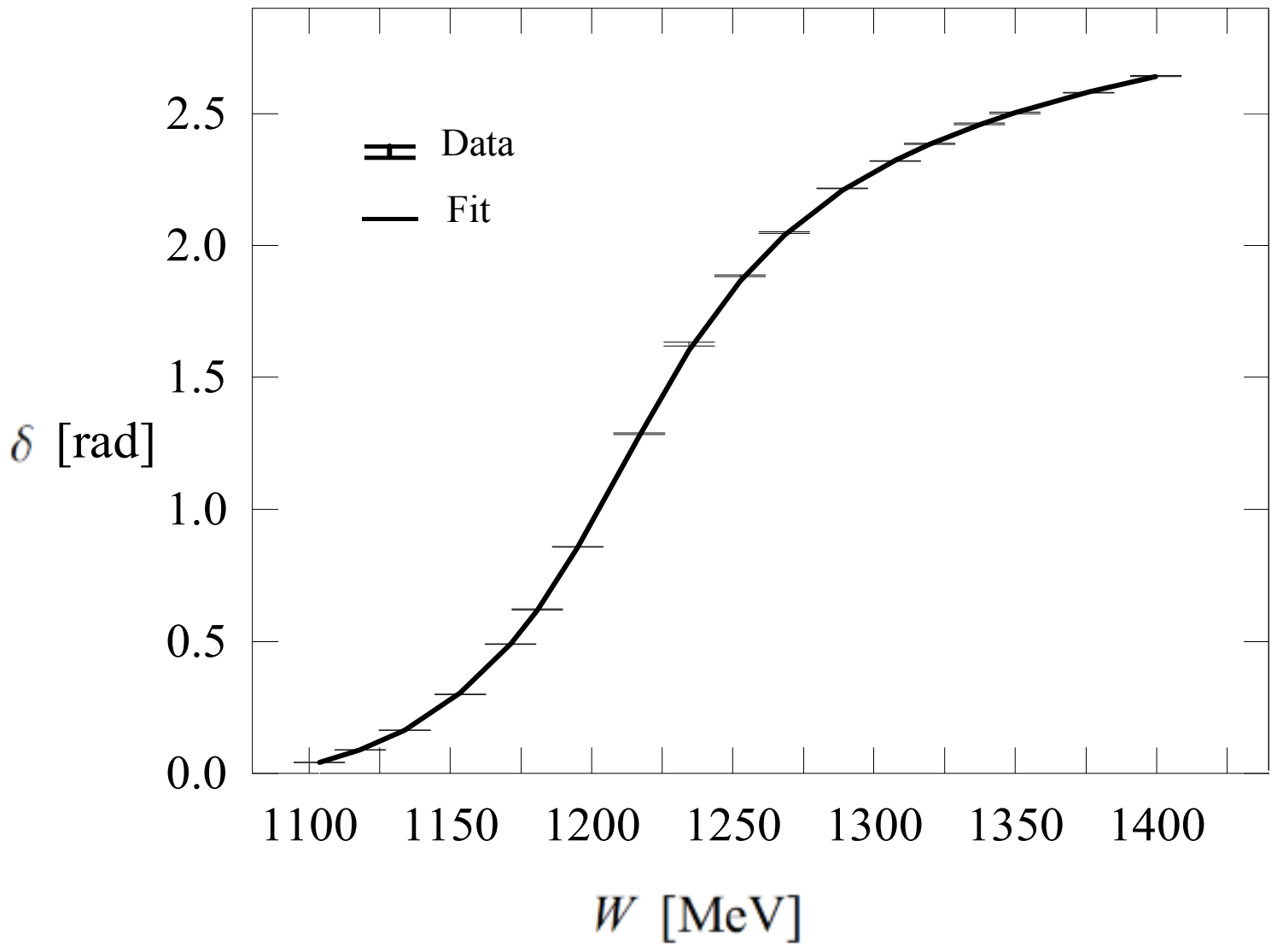

**Figure** 1. The Scattering phase shift $\delta$ for the partial wave $P_{33}$.

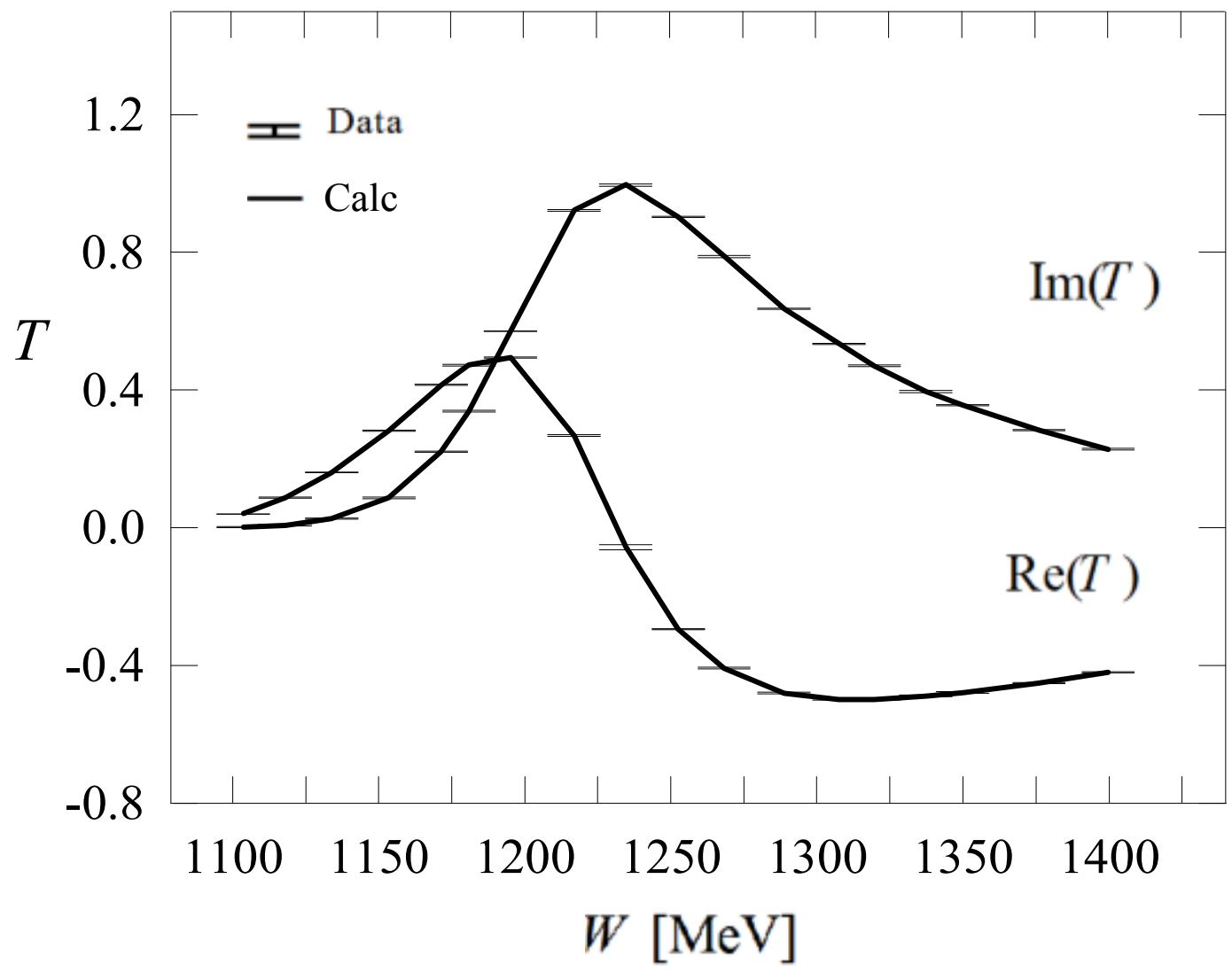

**Figure** 2. Real and imaginary parts of the transitional amplitude $T$.

In figure 3 we present the calculated time delay of the Δ(1232) resonance, using Eq. (3), where the resonance appears as distinct peak in $\Delta t(W)$ as a function of energy. The height of the peak and its width at the half maximum can be read from the figure and are found, with the help of Eq. (15),

$$W_r = 1211 \text{ MeV} \qquad \Gamma = 104 \text{ MeV}$$

respectively.

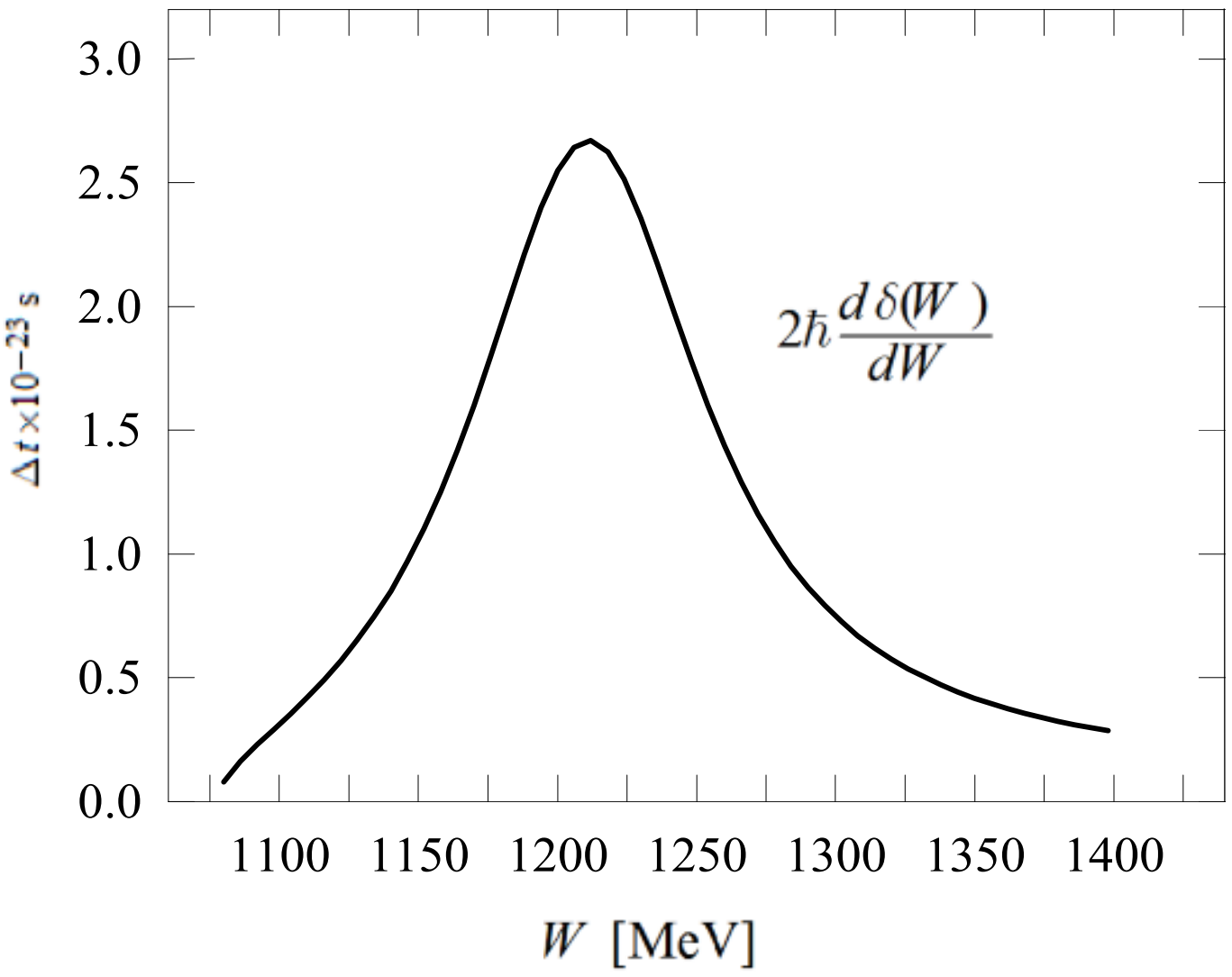

**Figure 3**. Time delay plot of the Δ(1232) resonance.

**Conclusion**

In this work we have employed our previously obtained parameterization of the scattering phase shift of $\pi N$ elastic scattering to evaluate the time delay plot. We used the time delay plot to extract the pole position and width of the Δ(1232) resonance for the partial wave $P_{33}$. The results of our calculations are determined from the graph to the best of our accuracy, and showed reasonable values which can be compared with other earlier determinations.